# Reversed Cherenkov-transition radiation in a waveguide partly filled with an anisotropic dispersive medium

T. Yu. Alekhina[a*] and A. V. Tyukhtin [a]

[a] *Saint Petersburg State University,*
*7-9 Universitetskaya Emb., St Petersburg 199034 Russia*

*E-mail:* t.alehina@spbu.ru

ABSTRACT: We analyze the electromagnetic field of a bunch that moves uniformly in a circular metal waveguide and crosses a boundary between a vacuum area and an area filled with an anisotropic dispersive nonmagnetic medium. The medium is characterized by the diagonal dielectric permittivity tensor with components possessing frequency dispersion of plasma types. The investigation of the waveguide mode components is performed with the methods of the complex variable function theory. It is shown that in compliance with the parameters of the medium, Cherenkov radiation (CR) generated in the filled area of the waveguide can have reversed direction in relation to the bunch motion. CR can penetrate into the vacuum area of the waveguide, that is, the intensive reversed Cherenkov-transition radiation (RCTR) can be generated. The main properties of this radiation are described, and essential differences from the RCTR in the case of isotropic left-handed medium are revealed.

---

[*] Corresponding author.

# Contents

## 1. Introduction

Reversed Cherenkov radiation (RCR) generated, in particular, in a left-handed material (LHM) which exhibits negative electrical permittivity and magnetic permeability, received recently much attention [1-3]. The energy flow density of this radiation makes an obtuse angle with the direction of the charge movement, as opposed to the ordinary (forward) Cherenkov radiation (CR). The RCR is interesting due to potential applications to detection of particles, diagnostics of particle bunches and generation of microwave and terahertz radiation.

     The LHMs predicted by Veselago [4] were extensively studied since the 1990s [5,6] when possibility of artificial realization of LHMs or so-called double-negative metamaterials (DNMs) was demonstrated. These materials are periodic structures of metal or dielectric elements with a small spatial period and, therefore, they can be described as some effective media. Next experimental studies have shown the realizations of DNMs in the GHz and THz frequency ranges [7,8]. The direct observation of the RCR in waveguide was reported, in particular, in [9]. Investigation of the charge radiation in different problems with an infinite and semi-infinite LHMs was performed in [10,11]. It was shown that so-called reversed Cherenkov-transition radiation (RCTR) is generated in the vacuum area as a result of the interaction of the RCR with the boundary. This radiation (as well as the RCR) propagates at an obtuse angle with respect to the charge motion direction.

     Note that we and other authors [12] use the term 'Cherenkov-transition radiation' (CTR) to underline the peculiarities of this phenomenon. The CTR is a radiation that occurs when CR passes through the medium boundary or reflects from it. It is obviously that such radiation is not CR "in the pure view". The CR field does not depend on time in the reference frame connected with a charge while the CTR field does not satisfy this condition. The CTR differs also from ordinary transition radiation (TR) which takes place at any boundary for arbitrary charge velocity and has continuous frequency spectrum even in a waveguide. The CTR has discrete frequency spectrum in a waveguide (the CTR frequencies coincide with the CR ones). Therefore, the CTR effect has the features of both Cherenkov radiation and transition one.

     The case of the waveguide partly filled with an isotropic LHM was considered in our work [13]. It was demonstrated that the RCTR can be the main part of the wave field in some domain

in the vacuum area of the waveguide, and the number of propagating modes (which compose the RCTR) is always finite. In particular, the RCTR can be monochromatic at certain conditions.

However, there are some difficulties in the investigation of radiation in LHMs connected with fabrication of isotropic DNMs and essential energy losses especially in THz and visible ranges. Thus, there is a need for finding other designs and ways to reduce the losses so that these novel materials will be more convenient for applications.

Analysis of the anisotropic DNMs in different situations was made in [14-16]. It was shown that the characteristics of electromagnetic wave propagation in such DNMs are significantly different from that in isotropic materials. Note that the main attention was only paid to dispersion equation and spectral density of energy, and electromagnetic field was not studied, but the structure of the electromagnetic field can be important especially in waveguides.

The situation with the semi-infinite nonmagnetic anisotropic medium with a plasma-type dispersion of the dielectric permittivity tensor was considered in [17]. For such a medium, the RCTR effect can also take place. This medium can be used instead of the LHM for generation of the RCR and the RCTR. Ordinary LHMs are usually fabricated from short metallic wires (responsible for dielectric permittivity) and rings (responsible for magnetic permeability) that assembled in a periodic cell structure. Note that magnetic response of the medium results in significant losses in THz frequency band [8]. Realization of the medium without rings can be easier than the LHM, and, moreover, one might expect that losses would be less than in the LHM.

In the present paper, we consider a waveguide with a partial filling with a nonmagnetic anisotropic dispersive medium of electrical type. We will also make a comparison between such problem and the problem with the isotropic LHM studied in [13].

## 2. General solution

We consider a bunch $q$ that moves with a constant velocity $\mathbf{V} = c\beta \vec{\mathbf{e}}_z$ ($c$ is the light velocity in vacuum) in a metal circular waveguide of the radius $a$ in a vacuum ($z < 0$) and flies into an anisotropic nonmagnetic dispersive medium ($z > 0$). The medium is characterized by the diagonal dielectric permittivity tensor with components

$$\varepsilon_{11} = \varepsilon_{22} = \varepsilon_r(\omega), \quad \varepsilon_{33} = \varepsilon_z(\omega), \quad \varepsilon_{r,z}(\omega) = 1 - \omega_{pr,z}^2 / (\omega^2 + 2i\omega\omega_d), \quad \varepsilon_{ij} = 0 \text{ for } i \neq j, \quad (2.1)$$

where $\omega_d$ is a parameter responsible for dissipation (we will set $\omega_d \to +0$ in final results). The bunch moves along the waveguide axis ($z$- axis) and crosses the boundary at moment $t = 0$. The bunch has some distribution along the $z$- axis and a negligible thickness, i. e., the charge density is $\rho = q\delta(r) \eta(\zeta)/(2\pi r)$, where $\int_{-\infty}^{\infty} \eta(\zeta) d\zeta = 1$, $\zeta = z - ct\beta$. Further, for definiteness and simplicity, we will consider Gaussian bunch with the distribution $\eta(\zeta) = \exp\left[-\zeta^2/(2\sigma^2)\right]/(\sigma\sqrt{2\pi})$, where $\sigma$ is a half-length of the bunch.

The strict solution of the problem is obtained traditionally as an expansion into a series of eigenfunctions of the transversal operator and satisfaction of the boundary conditions at the metal wall of the waveguide ($E_z = 0$ at $r = a$) and at the boundary between two media ($E_r$ and $H_\varphi$ are continues at $z = 0$). As in the case of two semi-infinite contacted medium [18], the field components can be written in form of a sum of 'forced' field ($q$) and 'free' one ($b$):

$$\mathbf{E}_{1,2} = \mathbf{E}_{1,2}^q + \mathbf{E}_{1,2}^b, \qquad \mathbf{H}_{1,2} = \mathbf{H}_{1,2}^q + \mathbf{H}_{1,2}^b, \qquad (2.2)$$

where subscripts 1 and 2 relate to the areas $z < 0$ and $z > 0$, respectively. The forced field is the field of a bunch in a regular waveguide.

Here we give only expression for the magnetic component of the forced field in the form of decomposition on normal modes [19]:

$$H_{\varphi 1,2}^q = \frac{2q\beta^2 c}{\pi a^3} \sum_{n=1}^{\infty} \chi_n \frac{J_1(\chi_n r/a)}{J_1^2(\chi_n)} \int_{-\infty}^{+\infty} \frac{\tilde{\eta}(\omega) h_{\varphi 1,2}^q \exp[i\omega\zeta/(c\beta)]}{\beta^2 c^2 k_{z1,2}^2 - \omega^2} d\omega, \qquad (2.3)$$

where

$$h_{\varphi 1}^q = 1, \quad h_{\varphi 2}^q = \varepsilon_r/\varepsilon_z, \quad k_{z1} = \sqrt{\omega^2 - \omega_n^2}/c, \quad k_{z2} = \sqrt{\varepsilon_r(\omega^2 - \varepsilon_z^{-1}\omega_n^2)}/c, \qquad (2.4)$$

$\omega_n = \chi_n c/a$, $\chi_n$ is the $n^{\text{th}}$ zero of the Bessel function $(J_0(\chi_n) = 0)$. Here, $\tilde{\eta}(\omega)$ is the Fourier transform of the longitudinal distribution $\eta(\zeta)$, normalized to Fourier transform of the distribution of a point charge: $\tilde{\eta}(\omega) = c\beta \int_{-\infty}^{\infty} \eta(-c\beta\tau) e^{i\omega\tau} d\tau$. The considered Gaussian bunch is characterized by the following real spectrum: $\tilde{\eta}(\omega) = \exp[-\omega^2 \sigma^2/(2\beta^2 c^2)]$. The forced field (2.3) contains CR (under certain condition).

The second summands in (2.2) which are the 'free' field components [18] connected with the influence of the boundary. The magnetic component is written in the form [20]

$$H_{\varphi 1,2}^b = 2q\beta/(\pi a^3) \sum_{n=1}^{\infty} \chi_n J_1(\chi_n r/a) J_1^{-2}(\chi_n) \int_{-\infty}^{+\infty} B_{n1,2} \tilde{\eta}(\omega) \exp[i(k_{z1,2}|z| - \omega t)] d\omega, \qquad (2.5)$$

where $B_{n1} = g(\omega)\left[(c\beta k_{z2} + \omega)^{-1} + \varepsilon_z(\varepsilon_r \omega - c\beta k_{z2})\varepsilon_r^{-1}(c^2\beta^2 k_{z1}^2 - \omega^2)^{-1}\right]$,

$B_{n2} = g(\omega)\left[\varepsilon_z(c\beta k_{z1} - \omega)^{-1} - (\omega + \varepsilon_r c\beta k_{z1})(c^2\beta^2 k_{z2}^2 - \omega^2)^{-1}\right]$, $g(\omega) = \varepsilon_r\left[\varepsilon_z(k_{z1} + \varepsilon_r k_{z2})\right]^{-1}$.

Note that $\text{sign} k_{z1} = \text{sign}\omega$ at $|\omega| \geq \omega_n$, and $\text{Im} k_{z2} > 0$ if we take into account some dissipation. This condition means that the waves outgoing from the boundary must decrease exponentially with an increase in distance $|z|$.

It should be noted that Eqs. (2.3) and (2.5) can be transformed to the known formulae for the case of semi-infinite medium [17] when we make the limiting process to the tube with the infinite radius $(a \to \infty)$. This limit transition can be performed using the Euler-Maclaurin formula (see, for example, [21]) which allows transforming a series to an integral.

Further, we analyse the exact solution for the field components of each mode (2.3) and (2.5) analytically (using the methods of the complex variable function theory) and numerically.

## 3. Analytical investigation

### 3.1 CR in the filled part of waveguide

First, we study the modes of the forced field (2.3) in the filled part of waveguide that is the field of a bunch in a regular waveguide with anisotropic filling under consideration. The condition for CR generation (i.e. the condition that the roots of the denominator of the integrand in (2.3) are real) is

$$\varepsilon_z\left(\beta^2-\varepsilon_r^{-1}\right)\geq 0. \qquad (3.1)$$

The most interesting situation takes place when $\varepsilon_z>0$ and $\varepsilon_r<0$ (or $\omega_{pz}<\omega_{pr}$) when CR is generated at any velocity of the bunch motion (see Eq. (3.1)). The CR frequencies are

$$\omega_{0n}=\omega_{pz}x_{0n},\quad x_{0n}=\left(S_n-\xi_n\right)^{1/2}\left(2-2\beta^2\right)^{-1/2}, \qquad (3.2)$$

where $S_n=\left[\xi_n^2+4\beta^2\left(1-\beta^2\right)x_a^2\left(y_n^2+1\right)\right]^{1/2}$, $\xi_n=\beta^2\left(y_n^2+x_s^2\right)-1$, $x_s^2=1+x_a^2$, $y_n=\omega_n/\omega_{pz}=\chi_n/x_p$, $x_p=\omega_{pz}a/c$. Parameter $x_a=\omega_{pr}/\omega_{pz}$ is responsible for the anisotropy of the medium (for the medium under consideration $x_a>1$). The frequencies of the radiated waves (3.2) lie into the range $\omega_{pz}<\omega_{0n}<\omega_{pr}$ ($1<x_{0n}<x_a$).

One can obtain [17] that, for this frequency range, the group velocities of the radiated waves in case of unbounded medium are directed at an obtuse angle with respect to the speed of the charge. So, such a medium exhibits properties similar to the properties of the left-handed medium. Note that for $x_a<1$ the group velocity is directed at an acute angle with respect to the bunch motion. The example of the medium with $x_a<1$ is strongly magnetised plasma where CR has more usual properties [23].

The forced field in the medium (2.3) consists of two parts: a quasi-Coulomb field and CR which has the magnetic component:

$$H_{\varphi 2}^{CR}=q/a^2\sum_{n=1}^{\infty}\chi_{n1}J\left(\chi_n r/a\right)J_1^{-2}\left(\chi_n\right)\tilde{\eta}\left(\omega_{0n}\right)A_n^{CR}\sin\left[\omega_{0n}\zeta/(c\beta)\right]\theta(-\zeta), \qquad (3.3)$$

where $\theta(x)$ is the Heaviside step function and the CR amplitude is

$$A_n^{CR}=4\beta^2\left(x_a^2-x_{0n}^2\right)/\left(x_p S_n x_{0n}\right). \qquad (3.4)$$

This expression is obtained traditionally with the residue theorem, as in the paper [20]. Note that $\tilde{\eta}\left(\omega_{0n}\right)$ is real value because of the symmetry of the bunch under consideration.

The behavior of the CR frequency (3.2) and amplitude (3.4) for the first mode of the wakefield for different values of $\beta$ and the anisotropy parameter $x_a$ are presented in Figure 1. CR is generated at any bunch velocity. At low velocities of $\beta$ the CR frequency is about $\omega_{pz}$ and the CR amplitude is relatively small. The frequency and the amplitude increase with an increase in the anisotropy and in the charge velocity.

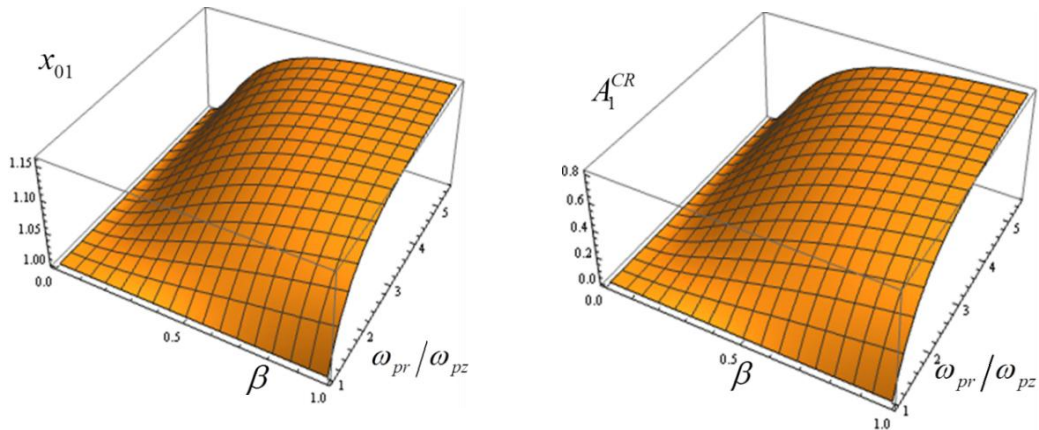

**Figure 1.** The behavior of the frequency (on the left) and the amplitude (on the right) of the first mode of CR at different parameters of the problem; nondimensional radius of the waveguide $x_p=4$, which corresponds $a=2.25$cm at $\omega_{pz}=2\pi\cdot 8.5$GHz.

## 3.2 The RCTR in the vacuum area of the waveguide

Asymptotic expressions for the free field components of each mode (2.5) are obtained with the steepest descent technique [24]. Note that the procedure used for the vacuum area ($z<0$) is similar to the procedure that was developed for different situations in our previous studies [10,11,13,17,20,22,23]. So, we only pay attention to the main points.

The first step in such research is the study of the singularities of integrands in (2.5) on a complex plane of $\omega$. One can show that integrands have the following singularities:
- the poles at the real axis $\pm\omega_{0n}$ (see Eq.(3.2)) and the poles in the imaginary axis $\pm\omega_{0n}^{(1)} = \pm i\beta\omega_n\left(1-\beta^2\right)^{-1/2}$ and $\pm\omega_{0n}^{(2)} = \pm i[S_n + \xi_n]^{1/2}\left(2-2\beta^2\right)^{-1/2}$;
- the branch points of the radicals $k_{z1}$ and $k_{z2}$ (2.4) $\pm\omega_n - i0$, and $\pm\omega_{pr} - i0$, $\pm\omega_{pz} - i0$, $\pm\tilde{\omega}_n^{(1)} = \pm\omega_{pz}x_s - i0$, correspondingly.

Figure 2 shows the singularities on the complex plane of $\omega$ and the comparative disposition of the integration pass. Note that the poles $\pm\omega_{0n}$ (3.2) and all of the branch points are slightly shifted downward from the real axis if small losses are taken into account. Therefore, the integration path goes along the real axis above these singularities and along the upper edge of the branch cuts defined by the equations $\text{Re}\sqrt{\omega^2 - \omega_n^2} = 0$, $\text{Re}\sqrt{\varepsilon_r\left(\omega^2 - \varepsilon_z^{-1}\omega_n^2\right)} = 0$.

In accordance with the steepest descent technique described in detail in [22] the poles and the branch points can be crossed at the transformation of the integration path into a new contour passing through the saddle points, and the contributions of the corresponding singularities should be included in asymptotic expressions. The saddle point contributions give space transition radiation, the branch cut contributions give the fields that exist only near the boundary, and the contribution of the poles (3.2) gives the RCTR.

The most interesting effect concerns the situation when CR penetrates through the boundary into the vacuum area, that is, the RCTR is excited in the vacuum area. If $\omega_n < \omega_{0n}$ ($y_n < x_{0n}$), the pole contribution is a propagating wave (the volume RCTR is generated). It can be shown that the number of propagating modes (which compose the RCTR) is always finite. In particular, it is possible to generate the RCTR consisting of a single mode, i.e. monochromatic

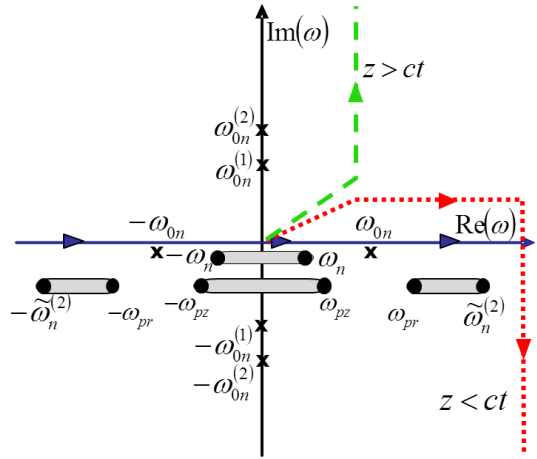

**Figure 2.** Disposition of the singularities of the integrands, branch cuts and integration paths (initial and transformed) in a complex plane of $\omega$ for the free field components in the vacuum area. The poles and the branch points are shown with crosses and circles, correspondingly.

RCTR. Note that, if $\omega_n > \omega_{0n}$ ($y_n > x_{0n}$), then this mode of the RCTR decreases exponentially with the distance from the boundary and does not transport the electromagnetic energy (CR totally reflected from the border).

Expression for the RCTR in the vacuum is the following:

$$H_{\varphi 1}^{CTR} = q/a^2 \sum_{n=1}^{\infty} \chi_n J_1(\chi_n r/a) J_1^{-2}(\chi_n) \tilde{\eta}(\omega_{0n}) A_n^{CTR} \sin\left[\omega_{0n}(t - \Psi_n z/c)\right] \theta(z_f - |z|),$$

$$A_n^{CTR} = 8\beta^2 x_{0n}\left(x_a^2 - x_{0n}^2\right)(x_p S_n)^{-1}\left[\Psi_n \beta\left(x_a^2 - x_{0n}^2\right) + x_{0n}^2\right]^{-1}, \quad \Psi_n = \sqrt{1 - y_n^2/x_{0n}^2}$$

(3.5)

The behavior of the amplitudes of the first mode of CR (3.4) and the RCTR (3.5) at different parameters of the problem are shown in Figures 1, 3. We can see that the behavior of the CR amplitude (Figure 1, right) differs from the CTR amplitude (Figure3, left) which has a maximum. If the bunch velocity is given then this maximum takes place at certain (not very large!) anisotropy of the medium. So, for a bunch with a velocity $\beta = 0.9$, this maximum is reached at $\omega_{pr} \approx 1.4 \omega_{pz}$. This peculiarity might be explained by the influence of the anisotropy on the reflection (and transmission) of CR from (through) the boundary. It is interesting that the RCTR amplitude can exceed the CR amplitude at certain parameters of the problem (Figure 3, right).

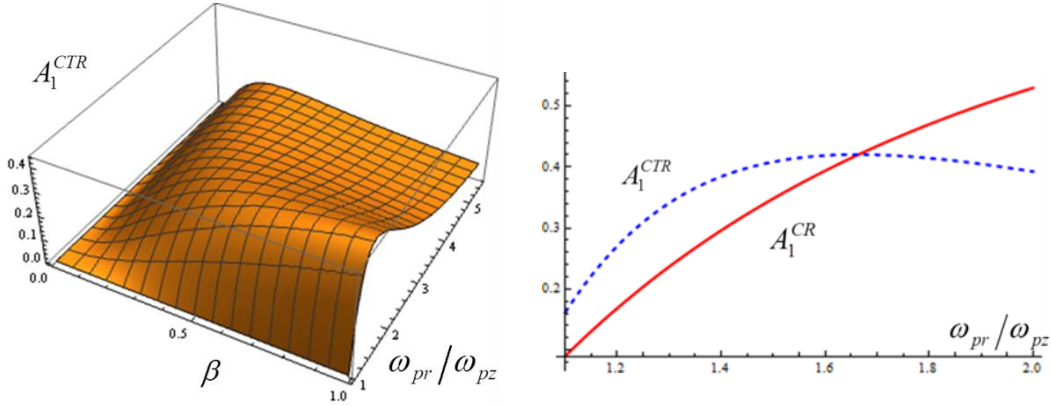

**Figure 3.** The behavior of the amplitudes of the first mode of CR and the RCTR at different parameters of the problem, $x_p = 4$; $\beta = 0.9$ (on the right).

The first mode of the total field component $H_\varphi$ in the vacuum area and in the medium is presented in Figure 4. CR (3.3) in the medium and the analytical approximation for the RCTR (3.5) are also presented. The method of computation of the exact integral representations for the free-field components (2.5) is based on a certain transformation of the initial integration path in the complex plane of $\omega$ as well as it was earlier maid for different problems [13,17]. We transform the initial integration path into a new contour in the upper-half plane for $|z| > ct$ (before the 'wave front') and into another contour in the lower-half plane for $|z| < ct$ (after the 'wave front'). These new contours are presented in Figure 2 with green dashed line and red dotted line which bypass all of the singularities and subsequently go parallel to the steepest descent path. Note that this algorithm allows overcoming a very abrupt behavior of the power exponent in (2.5) which is getting smoothed along new contours.

As can be seen, it is possible to reach the situation when the first Cherenkov mode is transmitted through the boundary and other Cherenkov modes are totally reflected (the case of the single-mode RCTR in the vacuum area). This effect takes place at $\chi_1 = 2.405 < x_p < \chi_2 = 5.52$. In the vacuum area, the free radiation field includes both the RCTR and the TR (Fig.4 shows the total field). The front of the RCTR propagates with the group velocity, and the RCTR exists in the domain $z < z_f = ct\,\Psi_n$. This inequality is obtained from the condition of the intersection of the pole at the transformation of the initial integration path to the steepest descent path (analogous procedure was described in [21]). One can see that the RCTR is the main part of the field in the vacuum area as well as CR is the main part of the total field in the filled area of the waveguide (Fig.4).

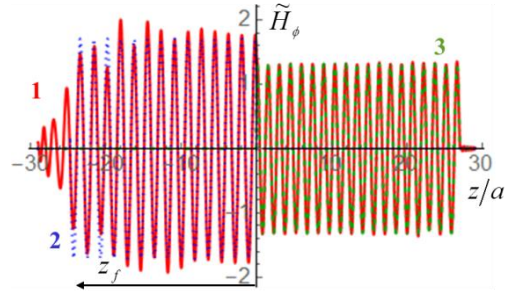

**Figure 4.** Dependence of the normalized transversal component $\tilde{H}_\phi = H_\phi a^2 / q$ of the 1$^{st}$ mode of the total field (solid red line **1**) on the distance $z/a$ at $r = 0.5$, $ct/a = 30$, $\sigma/a = 0.02$ and the velocitiy $\beta = 0.9$. The dashed blue line **2** pertains to the RCTR and the dot-dashed green line **3** pertains to CR; $n = 1$, $x_p = 4$, $x_a = 1.4$.

## 4. Comparison with the case of isotropic LHM

The RCTR effect takes place both in the medium under consideration and in the isotropic LHM considered in paper [13] where the medium is described by the following permittivity and permeability: $\varepsilon_2 = 1 - \omega_{pe}^2 \omega^{-2}$, $\mu_2(\omega) = 1 + \omega_{pm}^2 (\omega_r^2 - \omega^2)^{-1}$ (where $\omega_{pe}$, $\omega_{pm}$ and $\omega_r$ represent the electric and magnetic plasma frequencies and the resonant frequency, respectively). The main peculiarities of the phenomena in these media are presented in Table 1.

**Table 1.** The comparison between the anisotropic dispersive medium and the isotropic LHM.

| The anisotropic dispersive medium | The LHM |
|---|---|
| The CR generation range: $\omega_{pz} < \omega < \omega_{pr}$. | CR is emitted in the left-handed frequency range: $\omega_r < \omega < \min\left(\omega_{pe}, \omega_{sm} = \sqrt{\omega_r^2 + \omega_{pm}^2}\right)$ ($\omega_r < \omega_{pe}$) |
| CR in the waveguide has the reverse direction. ||
| The radiated frequency (3.2) $\omega_{0n} = \omega_{pz} x_{0n}$ | The radiated frequency $\omega_{0n} = \omega_r x_{0n}$ |
| The RCR is generated at any bunch velocity. ||
| If $\beta \ll 1$, then ||
| $x_{0n} = 1 + 2^{-1} y_n^2 (x_a^2 - 1) \beta^2 [1 + O(\beta^2)]$, $x_a = \omega_{pr}/\omega_{pz}$, $y_n = \chi_n / x_p$, $x_p = \omega_{pz} a/c$ $A_n^{CR} = 4\beta^2 x_p^{-1}(x_a^2 - 1)[1 + O(\beta^2)]$ | $x_{0n} = 1 + 2^{-1} x_{pm}^2 (x_{pe}^2 - 1) \beta^2 [1 + O(\beta^2)]$, $x_{pe,m} = \omega_{pe,m}/\omega_r$, $y_n = \chi_n / x_r$, $x_r = \omega_r a/c$ $A_n^{CR} = 4\beta^4 y_n x_{pm}^2 (x_{pe}^2 - 1)[1 + O(\beta^2)]$, |
| $A_n^{CTR} \approx 2 A_n^{CR}$ ||
| The RCTR can be the main part of the wave field in some domain in the vacuum area of the waveguide for certain conditions. ||
| The front of the RCTR propagates in vacuum with the group velocity. ||
| The number of propagating modes (which compose the RCTR) is always finite. ||
| If $y_n < 1$, then the RCTR mode with the number *n* exist in the vacuum area at any charge velocity. ||
| The first mode is the single one if ||
| $\chi_1 = 2.405 < x_p < \chi_2 = 5.52$. | $\chi_1 = 2.405 < x_r < \chi_2 = 5.52$. |
| If $1 < y_n < x_{0n}$, the n-*th* mode of the RCTR exist in the vacuum area at the velocity $\beta_n < \beta < 1$, ||
| $\beta_n = y_n (y_n^2 - 1)^{1/2} [x_a^2 - y_n^2 (y_n^2 + 1)]^{-1/2}$ | $\beta_n = y_n (y_n^2 - 1)^{1/2} [x_{pe}^2 x_{sm}^2 - y_n^2 (x_{pe}^2 + x_{pe}^2)]^{-1/2}$ |

## 5. Conclusion

We have shown that Cherenkov radiation generated in the filled part of the waveguide has the reverse direction and can penetrate into the vacuum area. The RCTR can be the main part of the wave field in the vacuum area. Also, it can be monochromatic for certain conditions. The RCTR mode amplitudes can exceed the CR mode amplitudes at some parameters of the problem (the bunch velocity and the anisotropy parameter). The maximum of the RCTR amplitude is reached at certain (not very large) anisotropy of the medium.

For the nonrelativistic case, the radiation is more intensive in the situation under consideration (the amplitudes are $\sim \beta^2$) in comparison with the case of the isotropic LHM (the amplitudes are $\sim \beta^4$). Note as well that the anisotropic medium under consideration can be simpler for fabrication and losses might be less in comparison with the isotropic LHM.

## Acknowledgments

This research was supported by the Russian Science Foundation (Grant No. 18-72-10137).